\documentclass[12pt]{article}
\usepackage{epsfig}

\parskip=\medskipamount 
plus.3em minus.5em \abovedisplayshortskip=.5em plus.2em minus.4em
\renewcommand{\thesection}{\arabic{section}}

\catcode`@=11

\@addtoreset{equation}{section} \@addtoreset{equation}{subsection}
\def\theequation{\ifnum\value{section}=0 \arabic{equation}\ignorespaces
\else \ifnum\value{section}=-1 A.\arabic{equation}\ignorespaces
\else \ifnum\value{subsection}=0
\thesection.\arabic{equation}\ignorespaces \else
\thesection.\arabic{subsection}.\arabic{equation}\ignorespaces
                             \fi
                        \fi
                   \fi}

{\catcode`\'=\active \def'{{}^\bgroup\prim@s}}

\catcode`@=12

\newcommand{\bq}{\begin{equation}}
\newcommand{\be}{\begin{equation}}
\newcommand{\fq}{\end{equation}}
\newcommand{\ee}{\end{equation}}
\newcommand{\bqr}{\begin{eqnarray}}
\newcommand{\beqs}{\begin{eqnarray}}
\newcommand{\fqr}{\end{eqnarray}}
\newcommand{\eeqs}{\end{eqnarray}}

\newcommand{\rf}[1]{(\ref{#1})}

\newcommand{\Dslash}{D\!\!\!\!/}

\def\bop#1{\setbox0=\hbox{$#1M$}\mkern1.5mu
    \vbox{\hrule height0pt depth.04\ht0
    \hbox{\vrule width.04\ht0 height.9\ht0 \kern.9\ht0
    \vrule width.04\ht0}\hrule height.04\ht0}\mkern1.5mu}
\def\Box{{\mathpalette\bop{}}}                        

\def\Prod{\prod}
  
\def\Triangle{\Delta} 

\begin{document}
\thispagestyle{empty}

\begin{flushright}
\begin{tabular}{l}
hep-th/0503110 \\
\end{tabular}
\end{flushright}

\vskip .6in
\begin{center}

{\bf  Masses and Interactions in Quantum Chromodynamics}

\vskip .6in

{\bf Gordon Chalmers}
\\[5mm]

{e-mail: gordon@quartz.shango.com}

\vskip .5in minus .2in

{\bf Abstract}

\end{center}

Correlations of composites corresponding to baryons and mesons are composed within 
the derivative expansion.  The expansion in energy scales permits a quantitative, 
algebraic description at various energy scales in QCD.  The masses in QCD are 
derived utilizing a proposed line interaction, with explicit checks of the masses 
up to the baryonic decuplet.

\vfill\break 

\section{Introduction} 

Correlator calculations in quantum chromodynamics are difficult due to the 
complicated nature of the diagrammatic expansion.  The derivative expansion has been 
recently developed to simplify these calculations, and in particular, to reduce the 
complicated integrals to a set of almost free-field ones.  Expansions pertaining to 
colliders are naturally formulated in terms of energies, from lower to higher ones; 
the derivative expansion is in this spirit.  The expansion is equivalent to the usual 
infinite number of loop graphs, but with the small parameter being the dimensionless 
ratio of energy scales as opposed to a coupling constant that could be of order unity.  

One of the outstanding questions in quantum chromodynamics, and generally in any gauge 
theory, is the origin of the masses of the gauge invariant states - in particular without 
resorting to computationally intensive lattice gauge theory.  In this work, in addition 
to formulating the correlator expansions, we derive the set of masses of the mesonic and 
baryonic tower of QCD masses, using the derivative expansion together with the inclusion 
of Wilson-like line integrals.  The latter operators are suited quite naturally within 
the derivative expansion.  

The masses of the composites containing the (s,d,u) multiplet, i.e. mesons and baryons 
and etc, are summarized here for convenience.  They follow via a formula, 

\bqr 
\langle {\cal O}(x_1) {\cal O}(x_2) \rangle = \sum_{j=1}^\infty c_j(x_1,x_2) 
 g^{2j} e^{-m\vert x_1-x_2\vert^{g^2}} \ .  
\label{twopointsum}
\fqr 
At $g\sim 1$, which is realistic for quantum chromodynamics, the 'propagator' or $2$-point 
function for the composites has the usual spatial dependence.  It is interesting that 
the coupling dependence of $g^2$ enters in this manner, which explains the differences 
in the masses via its flow in energies.  The lowest order approximation 
to these masses (containing the s,d, and u quarks) are found via the term in the 
exponential in \rf{twopointsum}.  It is calculated to contain two terms,  

\bqr 
m=\sum_\psi^{(j)} + {\tilde f} \ . 
\label{massterms}
\fqr 
The terms are due to the Wilson-like interaction,  

\bqr 
{\tilde f} = 150 \bigl( 4I -{2\over 3}\bigr) 10^6 \quad {\rm eV} \ , 
\fqr 
together with the individual fermionic mass terms in the (s,d,u) at the QCD scale, 

\bqr 
\sum m_\psi^{(j)} = 150 (N_\psi+N_S) 10^6 \quad {\rm eV} \ .
\fqr 
There are perturbative corrections to this formula via the power series $C_{j}(x_1,x_2)$ 
in \rf{twopointsum}, but the zeroth order formula in \rf{massterms} agrees well with the 
known mesonic and baryonic composites.  

This work is contained in the context of quantum field theory, and its placement in the 
context of supergravity is straightforward.  In previous work the mass generation of the 
fermion species content has been explained via gauge and gravitational instantons 
\cite{Chalmers5}, and when combined with the work here lends a fundamental explanation 
of the generation of the masses of the physical QCD sector.  Also, the gauge theory work 
presented in the current text may be generalized to finite temperature and supersymmetric 
field theories, but is not included.

\section{Brief review of QCD derivative expansion} 

Derivative expansions of quantum field and string theories have recently been developed 
wth several goals, one of which is to determine analytically their nonperturbative 
properties \cite{Chalmers1}-\cite{Chalmers9}.  This expansion is identical to the usual diagrammatic 
expansion in loops, but with an expansion in momentum scales as opposed to couplings.  
As a result, this approach commutes with dualities in supersymmetric theories and is 
generally nonperturbative in couplings.  One facet of this approach is that all integrals 
may be performed, and theories treated in this expansion have amplitudes that may be 
determined by a set of algebraic recursive equations, which are almost matrix-like.  
Gauge theories have been examined briefly in \cite{Chalmers2} and \cite{Chalmers3} in this 
context; we review the description of microscopic correlations describing amplitudes and 
composite correlations modeling nucleon interactions.  (These correlators are quantitatively 
related, however, for clarity we describe both.) 

The Lagrangian considered is 

\bqr
{\cal L} = \int d^4x \left( -{1\over 4} F^2 + \psi^a {\Dslash} \psi_a \right)
\fqr
quantum chromodynamics; the non-perturbative properties via coherent state Wilson 
loops and instantons are also examined.  The effective theory, expanded in derivatives 
is found from all possible combinations of gauge invariant operators ${\cal O}^{(j)}(x)$, 

\bqr 
{\cal S} = \int d^4x \sum_{j=1}^\infty h_j(g,\theta) {\cal O}^{(j)}(x)  
\label{operatorset} 
\fqr 
and $h_j(g,\theta)$ contains the full coupling dependence.  Example gauge invariant operators 
are ${\rm Tr} F^2{1\over\Box} F^2$, and $1\over m_\psi^2\psi^4$.  In the derivative expansion, 
self-consistency of the effective action with unitarity, implemented via sewing, allows for a 
determination of the functions $h_j$.  The action is next examined with respect to both 
logarithmic modifications of the terms and regulator dependencies.  

In addition to the hard dimension labeling the operator, logarithms also in general modify 
the form of the generating function, through, for example, 

\bqr  
{\rm Tr} F^2 \left[ \ln^{n_1}(\Box) \ln^{n_2}(\Box) \ldots \ln^{n_m}(\Box) \right]  
  \Box^2 F^2 \ , 
\fqr 
with covariantized boxes.  The presence of logarithms is required by unitarity and are 
generic in loop integrals; there are generically L multiplicative log terms at loop order 
L in the loop expansion.  These terms may be computed either in a direct sense via their 
inclusion in the effective theory, or may be determined by unitarity.  The logarithms are 
required via ${\cal I}{\rm m} S=S^\dagger S$ and may be computed from the analytic terms 
after their coefficients are determined.  

The form of the series expansion in terms of the operators depends on how the gauge field 
is regularized, in string theory with the string inspired regulator and dimensional 
reduction.  In the former there is a dimensional parameter $(\alpha')$ acting effectively 
as a cutoff; there may in general be other geometric scales depending on the model that 
may serve in the same role as $\alpha'$ in the following.  As the generating function 
contains one-particle reducible graphs, there must be inverse powers of derivatives, 
which are local in the sense that in momentum space these terms simply model the propagator 
$1/(k^2+m^2)=1/m^2\sum (-k^2/m^2)^n$; in the massless case the $1/\Box$ occurs, and in 
gauge theory their universal form, in the sense of independence of the number of external 
particles in a correlator, is expected based on collinear and soft factorizations.  Last, 
on-shell gauge theory amplitudes have infrared singularities; in the x-space expressions 
these singularities are absent as the lines are effectively off-shell.  

The regulator dependence of an $\alpha'$, or other dimensional parameter such as a mass 
term not depending on $\alpha'$ or a geometric parameter, follows in a straightforward sense 
by allowing their powers to occur in the expansion, i.e. $\alpha'^n m^{-p}$.  This occurs 
in quantum field theory via the decoupling of massive states, and in low-energy effective 
field theory as an expansion about an ultra-violet cutoff.  The general term in the effective 
action we consider is determined by including all gauge invariant terms discussed in the 
previous paragraph together with the these dimensional paramaters.  (In a dimensionally 
regularized theory the effective ultra-violet cutoff is absent and only the mass terms, 
with any other dimensionful parameters, occurs in the derivative expansion.)   Parameters 
such as Yukawa couplings in a spontaneously broken context occur in a polynomial 
sense as dictated by perturbation theory.  

The gauge coupling expansions of $h_j$ follow from the usual expansion of the gauge 
theory amplitudes, 

\bqr 
h_j(g,\theta) = \sum_{n=0}^\infty a_j^{(n)} g^{2+2n}  \ , 
\fqr 
and a series of non-perturbative terms, 

\bqr  
{\tilde h}_j(g,\theta)= \sum_{n=1}^\infty {\tilde a}_j^{(n)} e^{n(-{4\pi\over g^2}+i 
 {\theta\over 2\pi})} \ . 
\fqr 
The coefficients $a_j^{(n)}$ are determined via the sewing relations.  The instantons in 
the background field method generate ${\tilde a}$; potentially these contributions are 
redundant with the exponentiated gauge field integrals.  

For purposes of reviewing we formulate the four-point scattering of gauge bosons.  
The effective vertices to be inserted into the derivative diagrams are the interactions 
$(A,A^m\psi^k)$, $(A^2,A^{m-1}\psi^k)$, and $(A^3,A^{m-2}\psi^k$.  They are found by 
variation of the effective action, 

\bqr
v_1^{\mu;m,n}, \qquad v_2^{\mu\nu;m-1,n},  \qquad v_3^{\mu\nu\rho;m-2,n} \ .
\fqr
The unitarity relation that generates the full four-point amplitude function, in k-space, 
is, 

\bqr 
\int \Prod_{q=1}^{m+n} d^dq_j \quad v_1^{m,n}(k_1;q_i) \prod^m \Triangle_A 
 \prod^n \Triangle_\psi \quad v_3^{m-2,n} (q_i; k_2+k_3+k_4) 
\fqr 
\bqr 
+ v_2^{m-1,n}(k_1+k_2;q_i) \prod^m \Triangle_A \prod^n \Triangle_\psi v_2^{m-1,n}(q_i;k_3+k_4) 
\fqr 
\bqr  
+ {\rm perms} = v_4 (k_1,k_2,k_3,k_4) \ , 
\fqr 
in which the full derivative dependence has been implied in the vertices, and should 
be expanded termwise.  The integrals are easier to evaluate in $x$-space, as in \cite{Chalmers4}.  
The indices of the fields have been suppressed for 
notational purposes.  The propagators are indexed by $A$ and $\psi$.  In order to 
generate the full amplitude at the four-point, including the coefficients at general 
order $g^{2+L}$, we have to include the sewing relations that generate all of the other 
vertices; the infinite tower of unitarity relations are coupled and together generate 
the complete effective action corresponding to the loop expansion.  In general this 
appears complicated; however, to a finite order in coupling only a finite number of 
vertices are involved, which is small at low orders in coupling.  Furthermore, all of 
the integrals may be performed (including the massless ones using analytic methods 
as outlined in \cite{Chalmers3} and \cite{Chalmers4}).  If interested in computing to high 
orders in $g_{\rm YM}$ then the method is amenable via direct calculation or in a computer 
implementation.  

\begin{figure}
\begin{center}
\epsfxsize=12cm
\epsfysize=12cm
\epsfbox{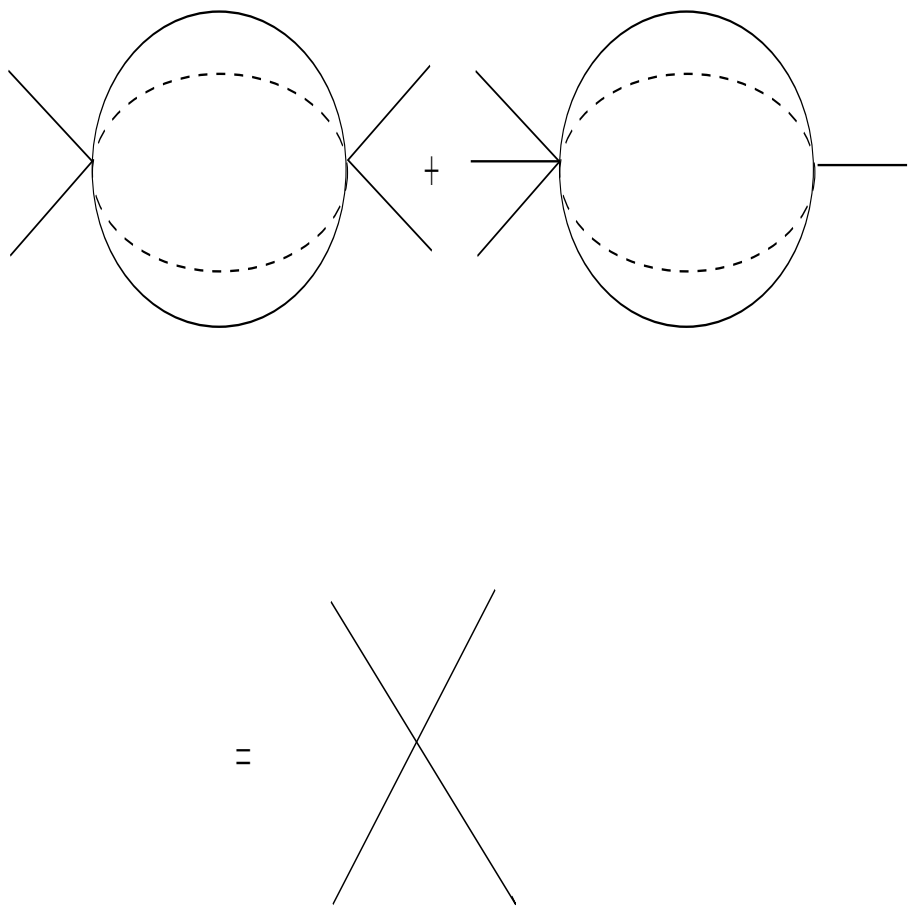}
\end{center}
\caption{The sewing relation illustrated at $4$-point.  Permutations are not included.}
\end{figure}

An explicit evaluation of the terms and integrals have been performed in massive 
scalar field theory, in an arbitrary dimension.  The reader is refered to \cite{Chalmers4} to 
see the simplest implementation.  

Next we examine the composite operator correlations; the composite operators ${\cal O}_j$ 
model the bare nucleons in terms of free particle states.  Flow of momentum amongst the 
various free-particle states in the composite operator is general.  A schematic is 
illustrated in figure 2.  In the correlations involving the composite operators, and in 
order to make contact with the parton model in perturbation theory, the internal lines 
of the operators (nucleons) are connected to a) full interaction vertices or b) from one 
nucleon to another.  In other words, the vertices (depicted in figure 4) are one particle 
reducible so that the perturbative contributions to the interactions in the usual loop 
expansion is obtained.  The interactions are depicted in figure 3 for a sample collision 
of three $\psi^3$ hadrons.  Figure 4 illustrates a usual interaction graph with that in 
the derivative expansion.  

\begin{figure}
\begin{center}
\epsfxsize=12cm
\epsfysize=12cm
\epsfbox{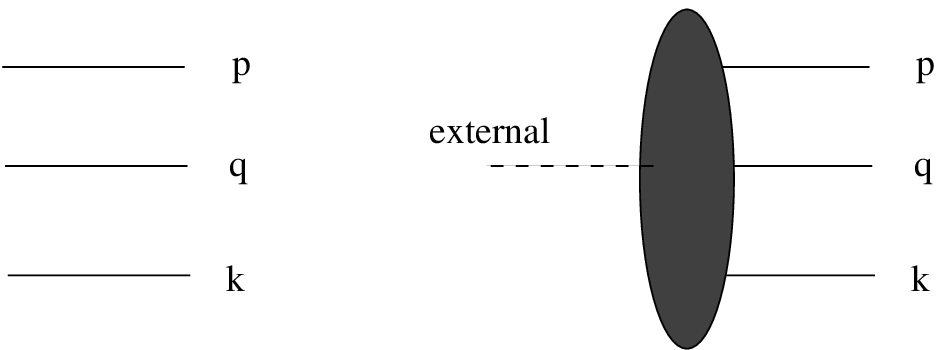}
\end{center}
\caption{Momentum flow of diagram and comparison with the usual parton picture.}
\end{figure}

\begin{figure}
\begin{center}
\epsfxsize=12cm
\epsfysize=12cm
\epsfbox{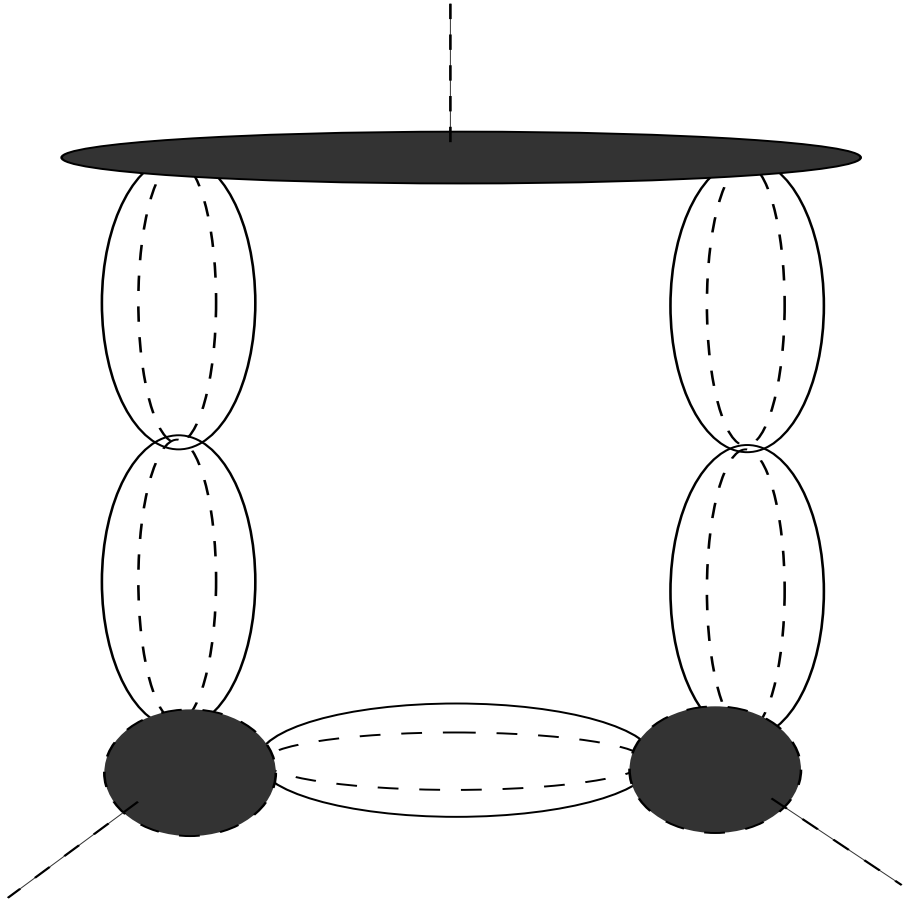}
\end{center}
\caption{A sample composite operator correlation corresponding to three baryon interaction.}
\end{figure}

\begin{figure}
\begin{center}
\epsfxsize=12cm
\epsfysize=12cm
\epsfbox{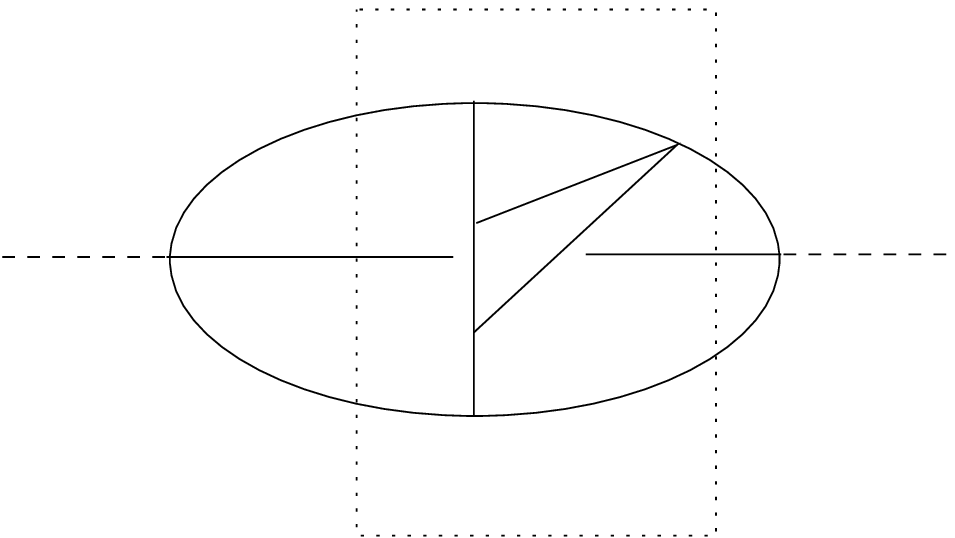}
\end{center}
\caption{A sample relation between the loop graph and derivative graph, found by expanding 
the integral.}
\end{figure}

\section{Exponential insertions and Masses} 

In this section integrals along 1-cycles and 2-cycles are included in the expansion, via 
the gauge field $A$ and the curvature $F$.  In doing so the masses of the nucleons are 
derived.  Their inclusion in the correlations described in the previous section depart 
from the parton model in that the realistic masses of the nucleons are obtained.  First 
the exponentials are obtained, together with their interacting exponentiated relatives, 
mediated by the microscopic theory (interacting in gauge coupling $g$); both are depicted 
in Figure 5.  The exponentials resemble flux tubes.  

The exponential operator we consider is 

\bqr  
{1\over 2} e^{-\alpha \oint A} P_G \ , 
\fqr 
with end-points fixed at the location of the composite operators.  Another operator 
that may be considered involves a curvature term, 

\bqr 
e^{{\tilde\alpha}\oint F} {\tilde P}_G \ , 
\fqr 
with an integral taken over compact Riemann surfaces attached at the operator locations.  
The line integral we take to be oriented in accord with the action of the isospin operator, 
the projection $P_G$ is an isospin operator acting on the composite operators, with an 
explicit factor of a half inserted to agree with the orientation of the integral;  Its 
eigenvalue on the nucleon made up of $n$-fermions with maximal isospin $I$ is 

\bqr  
\lambda = 4I -{2\over 3} \ . 
\fqr 
The contraction of a product of free-particle Wilson lines between two points $x=x_1-x_2$, 
by dimensional grounds and finiteness of the integrals, 

\bqr  
\prod^n e^{-\alpha\oint A} \rightarrow e^{n{\alpha^2/2} \ln(x^2)} \ , 
\fqr 
with $\alpha$ a general coupling constant, taken for example as $g_{QCD}\sim 1$ near the 
QCD scale.  The summation over an arbitrary number of closed loops, without taking into 
account interactions between them is, 

\bqr 
\sum_{n=0}^\infty {(-\lambda)^n\over n!} e^{n{\alpha^2\over 2}\ln(x^2/\mu^2)} = 
 \sum_{n=0} {(-\lambda)^n\over n!} ({x^2\over \mu^2})^{n\alpha^2/2}  
 = e^{-\lambda(x^2/\mu^2)^{\alpha^2/2}} \ . 
\fqr 
The dimensional parameter $1/\mu^2$ is introduced as an effective size of the Wilson 
loop; its natural value at the QCD point is 150 MeV ($\mu=12.23$).  As $\alpha$ scales 
to zero, e.g. at high energy, these pseudo nonperturbative contributions vanish; as $\alpha$ 
approaches one the exponential has the form to model a mass term in a propagator.  The 
coupling in QCD is of order unity, and as a result these contributions naturally model 
the mass of a nucleon.  

A nucleon is not a fundamental particle, and as a result one does not expect a propagator 
in the sense of $\partial\partial\Triangle=-\delta$ to model its dynamics.  The free 
particle composite of $n$ fermions, as in the parton model, has the form $\Triangle^n(x)$, 
containing $exp(-xnm_\psi)$.  The mass of the nucleonic state is found by computing the 
correlation $\langle {\cal O}(x_1) {\cal O}(x_2)\rangle$, 

\bqr  
\langle {\cal O}(x_1) \left[ {\rm Tr} \prod e^{-\alpha\oint A} P_G\right] {\cal O}(x_2) 
\rangle \sim C(x) e^{-f(x^2/\mu^2)^{\alpha^2\over 2}} 
\fqr 

\bqr 
C(x) e^{-\sum m_\psi^j x} e^{-{\tilde f} x}  \qquad \alpha\sim 1; \qquad {\tilde f}=f/\mu \ . 
\fqr 

The projection operator is taken to act in both directions along the line integral of 
$\oint A$.  The variable $\tilde f$ is, 

\bqr 
{\tilde f} = 150 \bigl( 4I -{2\over 3}\bigr) 10^6 \quad {\rm eV} \ , 
\fqr 
and the mass sum for fermions in the (s,d,u) at the QCD scale is approximately, 

\bqr 
\sum m_\psi^j = 150 (N_\psi+N_S) 10^6 \quad {\rm eV} \ . 
\fqr 
One can check that the mass formula agrees quite well with the masses in the baryon 
octet and decuplet, and the meson vector nonet (the bare masses of the fermions are 
taken as approximately 150 MeV and 300 MeV for the (u,d) and s quarks at the QCD scale.)  
This approximation is in the free-field point of view, and resummations of gauge 
interactions could modify the 'mass' of the nucleon - in quotes because an interacting 
nucleon is not really a particle.  The meson octet does not nearly agree as well as the 
rest; possibly this is due to the odd parity of these states and electroweak interactions.  
If the quarks were massless, then a $2$-quark state is differentiated from $n>2$ because 
of infra-red divergences, as can be seen by the first order gluonic correction to the 
two meson correlation.  Furthermore, it would be interesting to attempt to derive the 
subtle mass difference between the proton and neutron, or other degenerate states; the 
perturbative corrections to the mass calculation are desired.  

To compare, we list the known meson and hadron masses for the first few multiplets.  The 
quark content and masses of the baryonic octet are, 

\bqr 
\pmatrix{ I_3 : & -1 & -{1\over 2} & 0 & {1\over 2} & 1 \cr 
N(939): & & n\quad udd & & p \quad uud & \cr
\Sigma(1193), \Lambda(1116): \Sigma^- \quad dds & & \Sigma^0,\Lambda \quad uds & & \Sigma^+ 
 \quad uus \cr 
\Theta(1318): & & \Theta^- \quad dss & & \Theta^0\quad uss & \cr 
} 
\fqr 
and the same for the baryonic decuplet, 

\bqr 
\pmatrix{ I_3 & -{3\over 2} & -1 & -{1\over 2} & 0 & {1\over 2} & 1 & {3\over 2} \cr 
\Triangle(1232): \Triangle^- \quad ddd & & \Triangle^0 \quad ddu & & \Triangle^+ \quad duu 
 & & \Triangle^{++} \quad uuu \cr 
\Sigma(1384): & & \Sigma^- \quad dds & & \Sigma^0 \quad dus & & \Sigma^+ \quad uus \cr 
\Theta(1533): & & & \Theta^- \quad dss & & \Theta^0 uss & & \cr 
\Omega(1672): & & & & \Omega^- \quad sss & & & } \ .
\fqr 
These masses agree very well with the generated mass formula.

\begin{figure}
\begin{center}
\epsfxsize=12cm
\epsfysize=12cm
\epsfbox{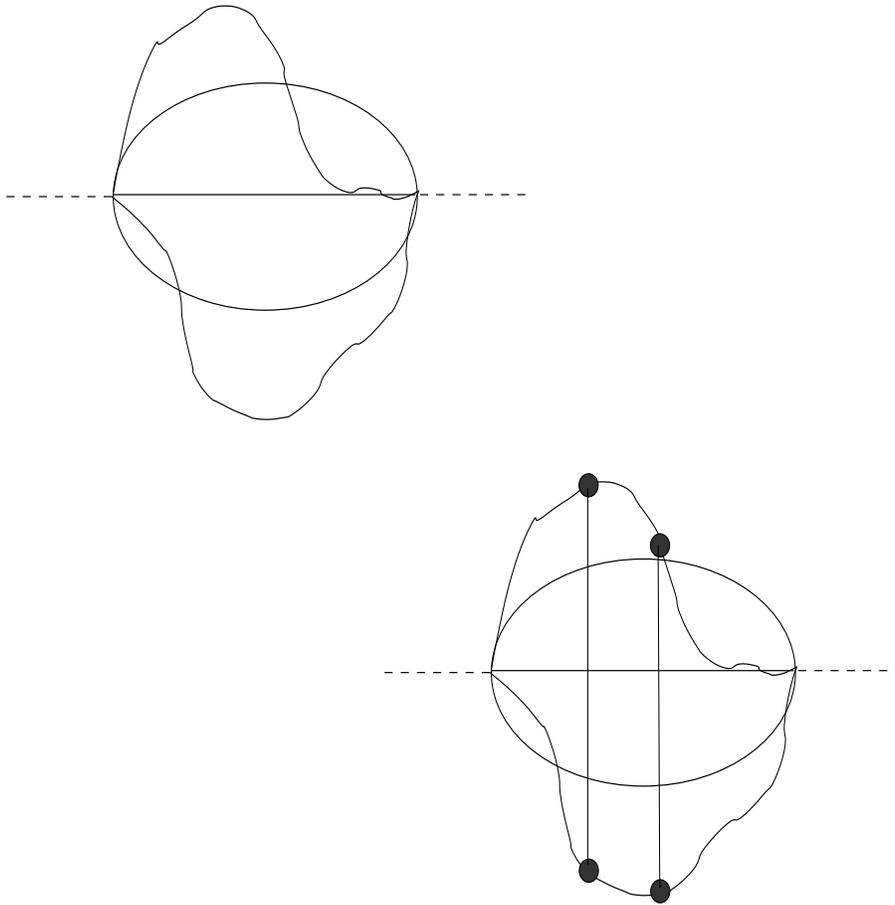}
\end{center}
\caption{The graphical interpretation of the $e^{-\alpha\oint A}$ inclusion.}
\end{figure}

The pseudoscalar mesons have masses: $\pi^\pm\quad (u{\bar d})$ 140 MeV, $\pi^0\quad (d{\bar d} 
- u{\bar u})/\sqrt{(2)}$ 135 MeV; $K^\pm(u{\bar s})$ 494 MeV, $K^0(d{\bar s})$ 498 MeV; 
$\eta_8\quad (d{\bar d}+u{\bar u} - 2s{\bar s})/\sqrt{(6)}$ 549 MeV; $\eta_0\quad (d{\bar d} 
+ u{\bar u}+s{\bar s})/\sqrt{(3)}$ 958 MeV.  The masses of the vector nonet multiplet are: 
$\rho$ 776 MeV; $K^\pm$ 892 MeV; $\omega(s{\bar s}$ 783 MeV; $\phi\quad (u{\bar u}+ d{\bar d}) 
/\sqrt{(s)}$ 1019 MeV.  The mass formula agrees well with the vector nonet and requires 
improvement with regards to the pseudoscalars.  

A primary aspect of the mass formula is that it exhibits approximate Reggeization.  The mass 
found in this approximations lay on $R^2$ parameterized by $I$ and fermion number $N_\psi+N_S$, 
illustrated in figure 6.  

The full 2-point correlator is obtained via inserting the vertex between the two free 
nucleonic states, and also summing the exponentiated gauge interactions, depicted in 
figure 7.  The vertex contracts any number of sets of lines, and there may be disconnected 
components connecting the lines between $x_1$ and $x_2$, as explained in section 2.  The 
correlator has a power series expansion, 

\bqr
\langle {\cal O}(x_1) {\cal O}(x_2) \rangle = \sum_{j=1}^\infty c_j(x_1,x_2)
g^{2j} e^{-m\vert x_1-x_2\vert^{g^2/2}} \ ,
\fqr
at $g\sim 1$, with corrections when the coupling is away from unity.  The vertex is 
obtained via the method in the previous section.  The potential soft dimensional terms 
in $c_j$ naively could resum to alter the mass term obtained from the interacting 
exponential gauge terms.  

\begin{figure}
\begin{center}
\epsfxsize=12cm
\epsfysize=12cm
\epsfbox{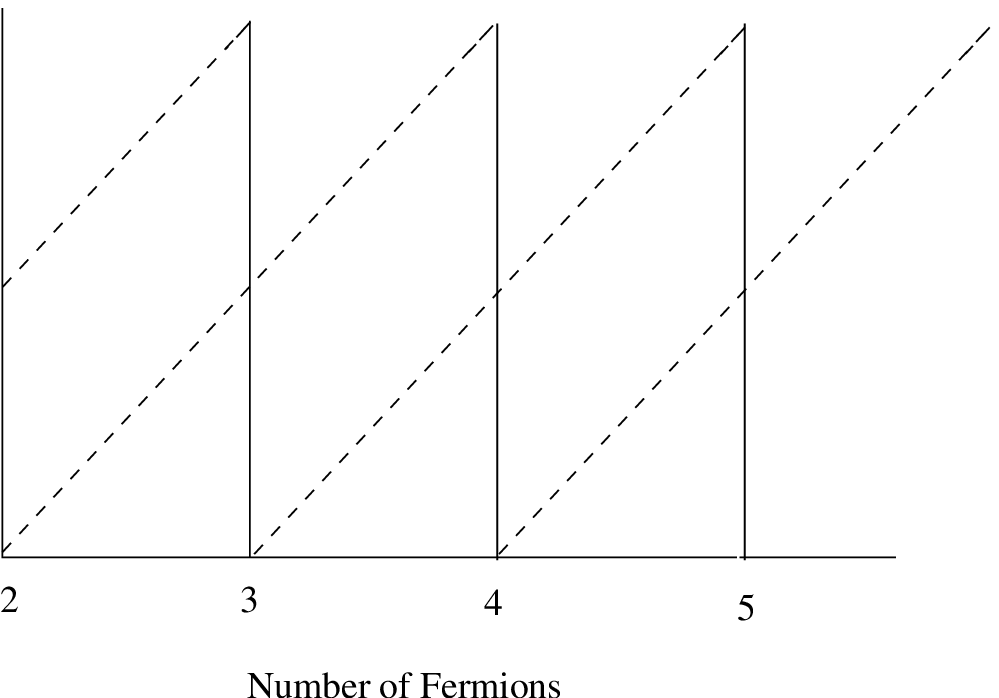}
\end{center}
\caption{Mass patterns as a function of the maximal isospin representation and fermion 
number.}
\end{figure}

\begin{figure}
\begin{center}
\epsfxsize=12cm
\epsfysize=12cm
\epsfbox{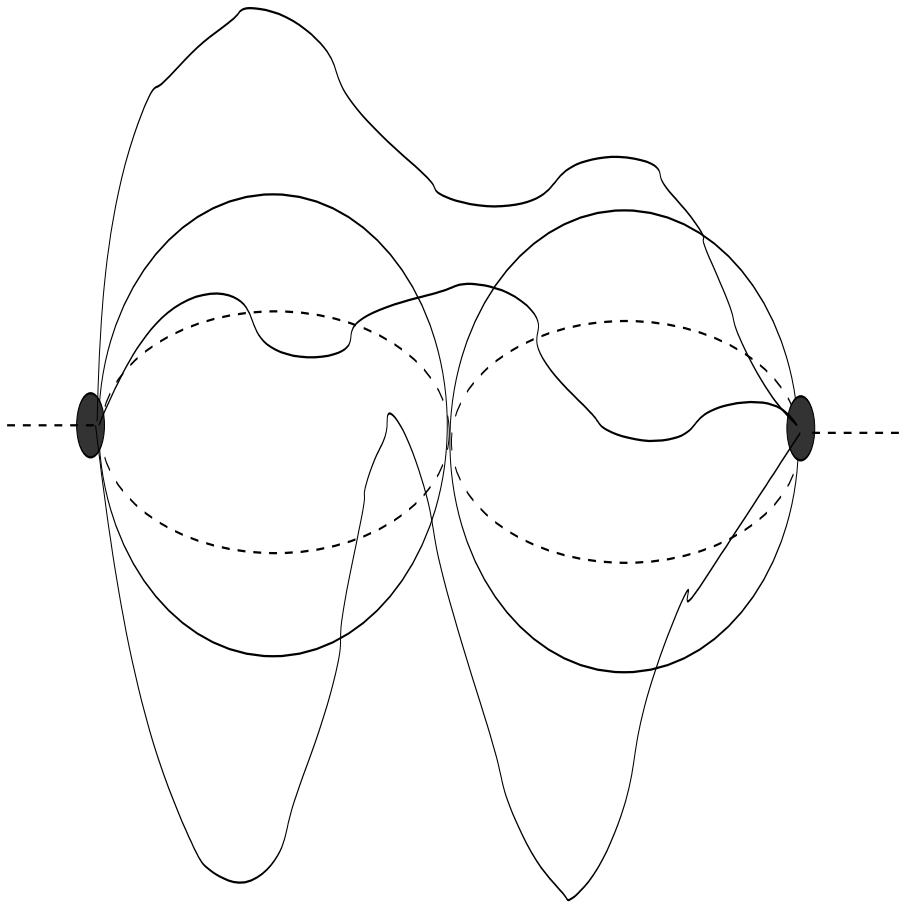}
\end{center}
\caption{Sample exponential terms contributing to the $2$-point correlator.  }
\end{figure}

\section{Nucleon interactions} 

The interactions follow via the interactions of the composite operators as described in 
section 2; these reproduce on a microscopic setting the parton model.  The additional 
interactions to be included are those of the Wilson loops.  Contrasted with the simplest 
two operator correlations, these exponentiated paths may join points $x_j$ in a number 
of ways for a given number of exponentials.  As a nucleon is not a point particle, the 
latter interactions model an effective mass of the constituent composite nucleons.  

\section{Discussion of general gauge theory} 

Two primary differences between QCD and a general gauge theoy are: the flow of the 
coupling constant and the energy scale of the theory.  These two properties change the 
mass formula and the scattering of the composite states.  For example, the mass formula 
has a $x^g$ in the exponential and depending on the UV properties of the theory (from the 
microscopic theory) the properties of the two-point correlator change; as usual, the 
composites may break into free particle constituents with an effective mass containing 
only the bare fermions.  The flow equations of the coupling constant dictate the fixed 
points of the theory.

\section{Conclusion} 

The correlations of nucleonic states are examined within the derivative expansion; these 
correlations are identical to those of quantum chromodynamics.  Additional exponential 
interactions, i.e. Wilson loops, are added to the interactions.  Masses of the nucleons 
are derived, and excellent agreement is found with the observed parameters.  They have 
a non-trivial dependence on the coupling constant, and in an asymptotically free limit, 
degenerate into the bare masses of the quark content without the gluon contribution.  
Small differences in the masses at given strangeness number within the multiplets are 
potentially found via perturbative corrections, for example, the mass splittings between 
the neutron and proton as well as other sets.  The fundamental masses of the quarks have 
also been analyzed and derived in the context of M theory via gravitational (and 
gauge) instantons.  

Explicit calculations and diagrams are presented that explain the interactions and 
methodology.  In general, gauge theories may be examined in the same approach;  the 
running of the coupling constant is governed by the perturbative expansion.  The 
derivation of the masses and their role in the dynamics may be found in the usual 
loop expansion.  Generalization to gauge theory with matter at finite temperature could 
be explained in the context presented in this work (with testable predictions for example 
at the RHIC collider).  

In the derivative expansion the calculations are much simplified over the usual 
perturbative loop approach.  The former is well suited to extend quantitative QCD work 
well into the lower energy regimes and to have a variety of quantitative applications, with 
latitude, at various energies.  

\vskip .3in 

{\bf Acknowledgements} 
\vskip .2in 

GC thanks Iouri Chepelev for comments.

\vfill\break

\vfill\break


\vskip .6in
\begin{center}

{\bf  Accurate Determination of Hadron and Baryon Masses}

\vskip .6in

{\bf Gordon Chalmers}
\\[5mm]

{e-mail: gordon@quartz.shango.com}

\vskip .5in minus .2in

{\bf Abstract}

\end{center}

In previous work a mass formula was generated within the derivative 
expansion that models the masses of the gauge invariant composites, 
such as mesons, hadrons, and baryons.  A possible improvement based 
on uniform corrections is given in this work that models the masses to 
an approximate .1 percent.  

\vfill\break  

In \cite{ChalmersA} QCD was reformulated in the derivative expansion.  
In there, a mass formula was generated that models the masses of 
the gauge invariant composites, 

\bqr 
M=150(N_\psi+N_s)+150(4I-{2\over 3}) 
\label{primarymass}
\fqr 
in terms of $10^6$ eV.  The first term is found from the bare 
masses of the fermions, and the second term is modeled from a 
superposition of 'Wilson' lines.  There is a similar mass formula 
\cite{ChalmersB} for the fundamental fermions, of the form, 

\bqr 
10^{n-4} \pm 2^m 5^i ~10^{-3} ~{\rm GeV} \ , 
\fqr 
or in MeV, 
\bqr 
10^{n-1} +/- 2^m 5^i ~{\rm MeV}  \qquad  n=1,2,\ldots 6  \ , 
\fqr 
which has a leading term plus a subleading term, which is 
quantized.  

It turns out that the addition of 

\bqr 
M'= 18 N \times 10^6 \quad {\rm eV}\ , 
\label{secondarymass}
\fqr 
to the formula in \rf{primarymass}, with $N$ a positive or negative integer 
models the masses to an approximate $.1$ percent.  The formula pertains 
to the hadronic and baryonic multiplets.  

The states and masses of the baryonic and hadronic multiplets are, 

\bqr  
\pmatrix{
 I_3 : & -1 & -{1\over 2} & 0 & {1\over 2} & 1 \cr 
N(939): & & n\quad udd & & p \quad uud & \cr 
\Sigma(1193), \Lambda(1116): & \Sigma^- \quad dds & & 
    \Sigma^0,\Lambda \quad uds & & \Sigma^+ 
 \quad uus \cr 
\Theta(1318): & & \Theta^- \quad dss & & \Theta^0\quad uss & \cr 
\pmatrix}
\fqr 
and the same for the baryonic decuplet, 

\bqr  
\pmatrix{ I_3 & -{3\over 2} & -1 & -{1\over 2} & 0 & {1\over 2} & 1 & {3\over 2} \cr 
\Triangle(1232): & \Triangle^- \quad ddd & & \Triangle^0 \quad ddu & & \Triangle^+ \quad duu 
 & & 
\Triangle^{++} \quad uuu \cr 
\Sigma(1384): & & \Sigma^- \quad dds & & \Sigma^0 \quad dus & & \Sigma^+ \quad uus \cr 
\Theta(1533): & & & \Theta^- \quad dss & & \Theta^0 \quad uss & & \cr 
\Omega(1672): & & & & \Omega^- \quad sss & & & \cr 
\pmatrix}  
\fqr 
These masses agree very well with the formula in \rf{primarymass}.  

The differences between the masses in \rf{primarymass} and the 'actual' 
masses are 

\bqr 
N(939) \qquad -18+7  
\fqr 
\bqr 
\Sigma(1193), \Lambda(1116) \qquad 5(18)+3,18-2 
\fqr 
\bqr 
\Theta(1318) \qquad 4(18)+2
\fqr 
and 

\bqr 
\Triangle(1232) \qquad -18  
\fqr 
\bqr 
\Sigma(1384) \qquad -18+2  
\fqr 
\bqr 
\Theta(1533) \qquad -18+1  
\fqr 
\bqr 
\Omega(1672) \qquad -18  \ .  
\fqr 
Some of the excited states also follow the same pattern, including, 
for example, 

\bqr 
\Xi(1321) \qquad 15(18)+1  
\fqr 
\bqr 
\Xi(1351) \qquad 15(18)+7 \ .
\fqr 
The corrections are normalized with the hadron masses.  

The subleading terms of $1,2,(7)$ are indicative of another uniform 
correction.  Perhaps 
the factors of $18$ and the prefactors are indicative of instanton-like 
or nonperturbative corrections.  If the factors of $18$ are indeed 
of the correct form, which is especially indicative in the baryonic 
multiplet, then the masses are found to an approximate $.1$ percent.  
The sub-leading terms of $18\times N$ appear to {\it justify} the formula 
in \rf{primarymass}, essentially in the baryonic decuplet, and the 
derivation in \cite{Chalmers1}.   

\vfill\break

\end{document}